\documentstyle[epsf,twocolumn,aps]{revtex}
\tightenlines

\newcommand{\be}{\begin{equation}}
\newcommand{\ee}{\end{equation}}
\newcommand{\ba}{\begin{array}}
\newcommand{\ea}{\end{array}}

\newcommand{\bac}{\begin{array}{c}}
\newcommand{\bal}{\begin{array}{l}}
\newcommand{\baR}{\begin{array}{r}}
\newcommand{\bacc}{\begin{array}{cc}}
\newcommand{\ball}{\begin{array}{ll}}
\newcommand{\balr}{\begin{array}{lr}}
\newcommand{\barl}{\begin{array}{rl}}
\newcommand{\baccc}{\begin{array}{ccc}}
\newcommand{\barcl}{\begin{array}{rcl}}
\newcommand{\balcl}{\begin{array}{lcl}}
\newcommand{\barcll}{\begin{array}{rcll}}
\newcommand{\barll}{\begin{array}{rll}}
\newcommand{\barrclcl}{\begin{array}{rrclcl}}
\newcommand{\bacl}{\begin{array}{cl}}
\newcommand{\bacll}{\begin{array}{cll}}
\newcommand{\eac}{\end{array}}
\newcommand{\ber}{\begin{eqnarray}}
\newcommand{\eer}{\end{eqnarray}}

\newcommand{\rueight}{RuSr$_2$GdCu$_2$O$_8$}

\newcommand{\rudelta}{RuSr$_2$GdCu$_2$O$_{8-\delta}$}

\begin{document}
\twocolumn[\hsize\textwidth\columnwidth\hsize\csname @twocolumnfalse\endcsname


\begin{center}
{\it Phys. Rev. Lett. (1999, in press).}
\end{center}

\title{Superconductivity in Ferromagnetic 
\rueight} 

\author{W. E. Pickett,$^{\dag}$ R. Weht,$^{\ddag}$ and A. B. Shick$^{\dag}$}

\address{$^{\dag}$Department of Physics, University of California, Davis CA 95616 \\
          $^{\ddag}$Departamento de Fisica, Comisi\'on Nacional de Energ\'ia At\'omica,
            1429 Buenos Aires, Argentina}

\maketitle

\tightenlines

\begin{abstract}
Relying on the inhomogeneous (layered) crystal, electronic, and magnetic
structure,
we show how superconductivity can coexist
with the ferromagnetic phase of
RuSr$_2$GdCu$_2$O$_8$ as observed by Tallon and coworkers.
Since the Cu $d_{x^2-y^2}$ orbitals couple only to apical O $p_x, p_y$
orbitals (and only weakly), which also couple only weakly to the 
magnetic Ru $t_{2g}$ orbitals, 
there is sufficiently weak exchange splitting, especially of
the symmetric CuO$_2$ bilayer Fermi surface, to allow singlet 
pairing.  The exchange splitting is calculated to be large
enough that the superconducting order parameter may be 
of the Fulde-Ferrell-Larkin-Ovchinnikov type.  
We also note that $\pi$-phase
formation is preferred by the magnetic characteristics of 
RuSr$_2$GdCu$_2$O$_8$.
\end{abstract}
\vskip 1cm
\footnoterule



\newpage

]

The antagonism between ferromagnetism (FM) and {\it singlet} superconductivity
(SC) was discussed early on by Ginsburg\cite{ginzburg}.
His simple conclusion, based upon an inverse Meissner effect that would set
up surface currents to shield the external region from the frozen-in
magnetic field $B_{int} = 4\pi M$, was that coexistence was not viable
except in samples not much larger that the field penetration depth.
Krey showed how to circumvent this restriction\cite{krey} 
by the formation of spiral magnetic order or,
in type II superconductors,
by the formation of a spontaneous vortex phase (SVP).  In the SVP the
internal magnetic induction is screened locally, vortex-by-vortex, so
the problem considered by Ginsburg does not apply.  Further work on SVPs
has included the suggested realization in ErRh$_4$B$_4$,\cite{kuper} in
Eu$_x$Sn$_{1-x}$Mo$_6$S$_8$,\cite{fischer} in ErNi$_2$B$_2$C,\cite{ng}
and possibly in p-wave systems.\cite{knig}

A serious impediment to SC arising well within the FM phase
is the Zeeman splitting of the carrier bands,
which makes the majority and minority Fermi surfaces inequivalent,
so the states $|\vec k\uparrow>$ and $|-\vec k\downarrow>$ do not 
both lie on the Fermi surface, and
total momentum $\vec q\equiv \vec k + \vec k^{\prime}$=0 
pairs are not available for
pairing.  Getting around this difficulty with $q\ne$0 pairs in the
case of applied fields or dilute magnetic impurities has led to
Fulde-Farrell-Larkin-Ovchinnikov
(FFLO) type theories,\cite{fflo} where either the SC or FM order
parameter (or both) develops spatial variation to 
accommodate the other.

Tallon {\it et al.}\cite{tallon,bernhard} have injected new 
excitement into this question of coexistence of SC and FM
by reporting the superconducting  
ferromagnet \rudelta~(Ru1212).  
This system was first reported by Bauernfiend {\it et al.}\cite{bauern}
as superconducting but not magnetic, and other reports\cite{felner,others}
indicate that properties are dependent on the method
of preparation.
Unlike almost all previously reported cases of
coexisting SC and FM, this material is first magnetic (T$_M$ = 132 K,
due to ordering of Ru ions with an ordered moment of 1 $\mu_B$/Ru)
and then becomes SC only well within the FM phase.
Superconductivity appears at T$_S \approx$ 35-40 K,
and only at 2.6 K do the Gd ions order (antiferromagnetically).
The data are reproducible, specific heat data indicate a bulk SC
transition, and
muon spin rotation experiments indicate the magnetism is homogeneous
and is unaffected by the onset of 
superconductivity.\cite{tallon,bernhard}  This
SC ferromagnet is quite different from previous materials\cite{maple}
where SC and FM order have similar
critical temperatures, compete strongly and
adjust to accommodate each other, and coexist only
in very limited regions
where magnetic order is small.\cite{feln1,feln2}

The observed phenomena present several interrelated questions.  The most
obvious is: how can SC exist with a FM material?  Secondly, how is the FM 
coupling transmitted between layers without killing 
superconductivity; 
T$_M$ = 132 K indicates electronic exchange coupling and not the much
weaker dipolar coupling.  Finally, how is the SC coupling propagated 
through the FM layers?  These are the questions that we address.

This hybrid ruthenocuprate Ru1212,
isostructural with insulating triple perovskite NbSr$_2$GdCu$_2$O$_8$,\cite{Nb} 
is comprised of double 
CuO$_2$ layers separated by a Gd layer, sandwiched in 
turn by SrO layers, as shown in Fig. 1.
The unit cell in completed by a RuO$_2$ layer, 
making it structurally
similar to YBa$_2$Cu$_3$O$_7$ except that the CuO chain layer is 
replaced by a RuO$_2$ square planar layer, with resulting 
tetragonal symmetry (except for distortions typical of perovskites).  

Magnetism is detrimental to superconductivity both through 
its coupling to spin and
to orbital motion, which we consider in turn.
Since there is a strong tendency for singlet pairing in materials
with CuO$_2$ layers such as Ru1212 has, and substitution of Zn for
Cu leads to a decrease in T$_S$ similar to that seen in cuprate SCs,
we examine specifically the possibility of SC 
CuO$_2$ layers.\cite{triplet}
There are three potential limiting mechanisms: (1) 
Zeeman splitting of pairs due to the dipolar
field $B_{int}$, (2) the electronically mediated
exchange field $\Delta_{ex}$
that also splits majority and minority Fermi surfaces, and
(3) charge coupling to the vector potential leading to supercurrents.
It is primarily the second item  
that presents difficulty for singlet SC in this system. 
We conclude that SC will most likely be accommodated by 
development of a FFLO-like modulation of the SC order parameter within
the CuO$_2$ layers, possibly accompanied by 
``$\pi$ phase'' formation.\cite{prokic}

\begin{figure}[tbp]
\epsfxsize=3.4cm\centerline{\epsffile{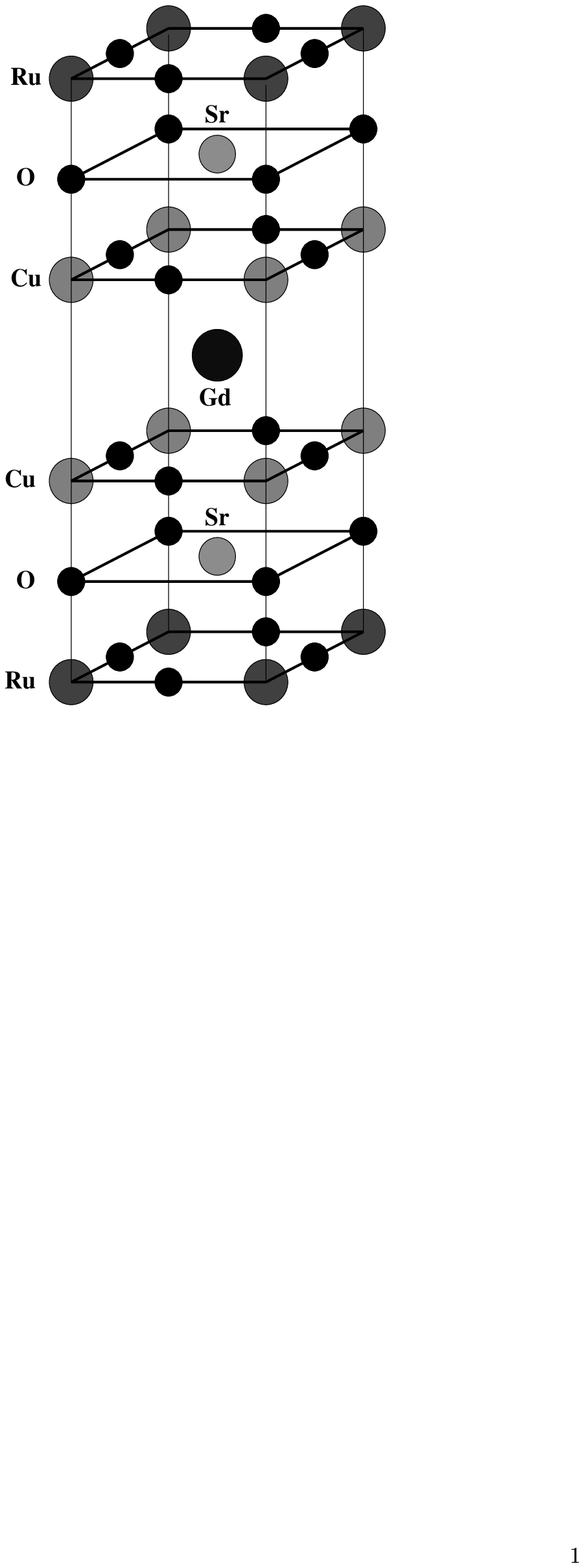}}
\caption{The crystal structure of RuSr$_2$GdCu$_2$O$_8$, with small
distortions of the RuO$_6$ octahedron and the CuO$_5$ pyramids
neglected.  Magnetism occurs in the RuO$_2$ layer, superconductivity
in the Cu-O bilayer.
\label{Fig1}}
\end{figure}

{\it Spin-derived Pair Breaking}.  It is easy to dispense 
with dipolar spin coupling [(1) above] 
due to the internal field.
The Ru magnetization corresponds to a macroscopic 
(volume average) field
induction $B_{int}=4\pi~<M>$=700 G, 
for which the Zeeman splitting 
(5 $\mu$eV) is negligible 
compared to the pair binding energy $2\Delta \sim$ 5 k$_B$T$_S$ $\sim$
15-20 meV as well as to the exchange splitting (discussed below).  
As mentioned in (2) above, the magnetization $M$ of the RuO$_2$
layer also gives rise to an induced
exchange field $B_{ex} \equiv 2\mu_B \Delta_{ex}$ 
in the CuO$_2$ layer that splits each 
CuO$_2$-derived Fermi surface (FS), with the larger (smaller) FS 
corresponding to the majority (minority) carriers.  
Unlike a real field, $B_{ex}$ couples only to the spin.

It is necessary first to obtain the magnitude and $\vec k$
dependence of this exchange splitting 
of the carriers in the CuO$_2$ layers.  To this end
we have applied density functional methods.\cite{wien}
Our calculations, using both the local density approximation (LDA)
and generalized gradient approximation (GGA),\cite{hts99} resulted in a
FM Ru-O layer as well as strongly spin
polarized Gd (moment of 7 $\mu_B$\cite{Gd} as expected).
The value of the moment in the Ru layer is sensitive to both the choice
of exchange-correlation functional (LDA or GGA) and also to structural
distortions, which will be discussed more fully
elsewhere.  Possible effects of correlation on the Ru moment were
checked by applying the LDA+U procedure\cite{sasha} with a Coulomb repulsion
U$_{Ru}$=3 eV.  The moment was very similar to the GGA value and
in all cases the RuO$_2$ layer remained metallic.
The calculated moment (using GGA) of 2.5 $\mu_B$ ($\sim 1 \mu_B$ lies on
the six neighboring O ions)
for the undistorted structure
is larger than the moment
of 1 $\mu_B$ reported by Tallon {\it et al.}  The sensitivity of the
calculated moment to oxygen positions suggests that using the true 
(distorted) crystal structure would reduce the discrepancy.
We regard our calculated 
exchange splitting in the CuO$_2$ bilayer as an upper bound
on the true value, which is sufficient for present purposes.
\begin{figure}[tbp]
\epsfxsize=4.6cm\centerline{\epsffile{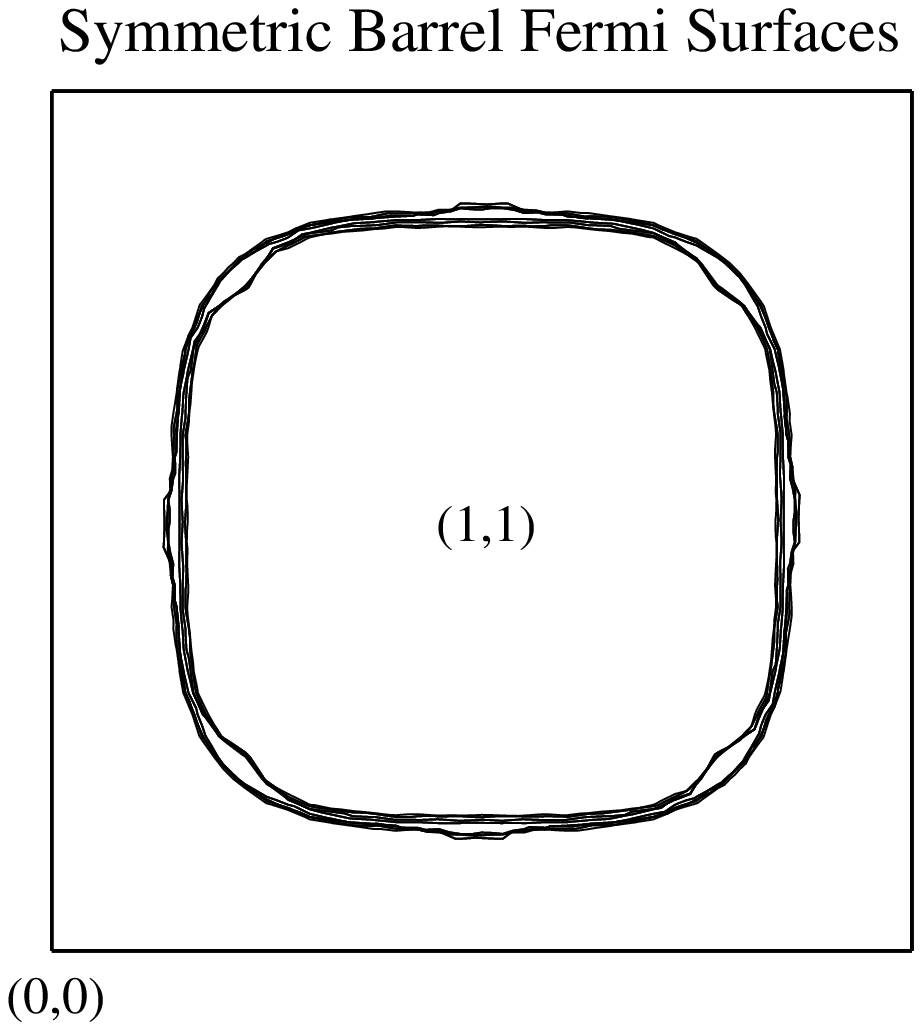}}
\caption{The symmetric CuO$_2$ barrel Fermi surfaces of RuSr$_2$GdCu$_2$O$_8$.  
Both majority and minority Fermi surfaces are shown, reflecting the
small spin splitting.  Coordinates shown are in 
units of $(\pi/a,\pi/a)$.
\label{Fig2}}
\end{figure}

As expected from previous theory and experiment,\cite{science} 
the CuO$_2$ bilayer
gives rise to two barrel Fermi surfaces (FS) of each
spin centered at the zone 
corner, with the inner (outer) FS corresponding to states 
(at $k_z$=0) that
are symmetric (antisymmetric) under the mirror operation 
connecting the two CuO$_2$ layers.
The symmetric FS, shown in Fig. 2, is regular in shape, has
a nearly $\vec k$-independent exchange splitting, and has
quasi-one-dimensional nesting features.
By direct close inspection of the
band structures, we obtain
the difference in Fermi wavevectors on this FS (not shown) 
$\delta k_F \sim 0.02 k_F$.  
The antisymmetric FS is less regularly shaped and, due to 
$\vec k$ dependent
hybridization with Ru, has $\vec k$ dependent exchange splitting that 
makes it less favorable for pairing.  Thus we concentrate
on the symmetric FS.

The small exchange splitting 
(compared to $\sim$1 eV in the Ru-O layer)
$\Delta_{ex} = v_F \delta k_F \approx$ 25 meV ($v_F = 2.5\times 10^7$
cm/s\cite{pba})
is a direct consequence of the electronic, 
magnetic and
crystal structure.  The Ru magnetization lies within the $t_{2g}$
orbitals, which couple with the apical O $p_x$ and $p_y$ orbitals
only through a small $pd\pi$ coupling.  These O $p\pi$ orbitals do not couple
either with the Cu $d_{x^2-y^2}$ orbital, which is the main character of 
the Cu-O barrel Fermi surfaces, nor do they couple with the Cu $s$
orbital, which has been found in YBa$_2$Cu$_3$O$_7$ to provide much
of the $\hat z$ axis coupling.  The exchange coupling that survives must
find a secondary route, such as through polarization of the apical
O atom that transfers the polarization to the $p_z$ orbitals and
on to the Cu $s$ orbital, or from the apical O to the O $p\sigma$
orbitals in the Cu-O layers.
This small exchange splitting can be regarded (for effects on
the spin) as arising from a
vector exchange field $\vec B_{ex}$, whose direction is 
linked to the direction $\hat M$ of
$\vec M$.  

{1. $\vec M$ parallel to the RuO$_2$ layers.}  In this case the vector
potential $\vec A$ can be chosen to be perpendicular to the layers.  Then
$\vec p \cdot \vec A$ orbital pair-breaking is 
confined to the interlayer hopping 
motion, which we neglect as suggested by Bernhard 
{\it et al.}\cite{bernhard}  The semiclassical Green's
function treatment of
Burkhardt and Rainer (BR)\cite{rainer} then applies, except that
the magnetic
field of that work is replaced by the effective exchange
field seen by the carriers
\begin{equation}
\vec B_{eff} = \vec H +\vec B_{int} + \vec B_{ex}
\end{equation}
comprised of all contributions to the spin splitting $\Delta_{Zee}
= 2 \mu_B |\vec B_{eff}|$ in the CuO$_2$
layers: an applied field $H$, the internal (dipolar) field
$B_{int}$ (equal to $4\pi M$ within the RuO$_2$ layer)
and the exchange field $B_{ex}$ induced in the CuO$_2$ layers
by the electronic exchange interaction ($B_{ex} \equiv 
\Delta_{ex}/2\mu_B$).  $B_{int}$ in the Cu-O bilayer is obtained from
magnetostatics or, below T$_S$, a generalized London equation.

BR have extended the FFLO theory, showing that in-plane ``fields'' 
$\Delta_{Zee} \geq 2 \Delta$ (the SC gap) can be
accommodated by a non-constant SC order parameter up to a 
maximum value $B_{c2}$.
Since the internal field $B_{int}$ due to 1 $\mu_B$/Ru is only
700 G,
for most of the range of accessible fields the exchange field $B_{ex}$
will be the limiting field.
In an FFLO state the mean pair momentum 
\begin{equation}
q = \delta k_F \approx 0.02 k_F \sim 0.02 \pi/a
\end{equation}
corresponds to a
SC order parameter modulation on the scale of $\lambda_q = 
2\pi/q \sim$ 400 \AA, which
must be no shorter than the 
SC in-plane coherence length $\xi_{ab}$.
For conventional cuprates  
with T$_S \sim 40$ K, for which $\xi_{ab} \sim$ 60-75~\AA, the exchange
splitting $\Delta_{ex}$= 25 meV (greater than 2$\Delta$) 
rules out a constant order
parameter but allows a non-constant SC
order parameter of a generalized FFLO type in the cuprate layers.
BR note that, while 
2D character enhances tendencies toward a FFLO-type state, the existence
of such a state can be sensitive to Fermi surface shape.  The quasi-1D
sides of the barrel FS (Fig. 2) should strongly favor an FFLO state.

{2. $\vec M$ perpendicular to the CuO$_2$ layer.}
For this orientation
coupling of orbital motion to the total field 
$\vec H + \vec B_{int}$
leads to supercurrents, 
and is naturally accommodated
in the superconducting CuO$_2$ bilayer as a SVL.
The lattice spacing corresponding to $M$=700 G (H=0)
is one flux quantum per circle of radius $\sim$0.7 $\mu$m,
posing no problem for coexistence.
At applied fields $H \gg B_{int}$, the effect of the intrinsic
magnetization becomes minor.  As a result, the Meissner effect 
measured in fields of a few Tesla may produce normal-looking susceptibility
curves, such as found by Tallon {\it et al.} (albeit on polycrystalline
samples).  The behavior of the susceptibility for $H\leq 4\pi<M>$ remains to
be elucidated.

{\it Interlayer Superconductive Coupling.}
Since bulk SC reflects a state that is 
coherent along the $c$ axis, pair-breaking by the
intermediate magnetic RuO$_2$ layer must not be so strong as to destroy
interlayer tunnelling of pairs (for which $\hat c$ axis hopping can
no longer be neglected).  Ru1212 represents the first atomic-scale
SC-FM superlattice, and although there exists a literature on 
{\it nanoscale} SC-FM superlattices, the theory has not been pushed
down to the atomic scale; indeed, no systems except
cuprates show superconductivity 
of a single atomic (bi)layer, which only becomes possible because the 
$\hat c$-axis coherence length $\xi_c$ is only $\sim$ 10\AA~(the
cell dimension).  

The present system is however 
a natural one to form the $\pi$-phase SC order parameter 
predicted for SC-FM superlattices.  The $\pi$-phase has an order
parameter that changes phase by $\pi$
from SC layer to SC layer, and thus has a node
in the FM layer, thereby
strongly decreasing the pair breaking effect.  Two characteristics of Ru1212
favor the $\pi$-phase.  First, the layer of strong 
magnetization is extremely thin
(the $\sim$ 2~\AA~of the RuO$_2$ layer).  Second, Proki\'c {\it et al.}
predict a $\pi$-phase only above a critical magnetization in the FM
layers, and the RuO$_2$ layer presents a rather high (RuO$_2$
layer) value of $4\pi M
\sim$ 4 kG within this atomic layer.  (The 700 G value mentioned above
is a cell average.)  Since the SC coupling strength in the (CuO$_2$)$_2$
bilayer is not known (and there is not theory of cuprate SC anyway)
a quantitative determination is not possible, but Ru1212 presents
a favorable case for $\pi$-phase formation.

{\it FM Coupling Through the SC Layers}.
We comment briefly on the FM order.  Since the magnetic ordering temperature 
depends only logarithmically\cite{imry} 
on the perpendicular coupling J$_{\perp}$,
the rather high Curie temperature is not inconsistent with the small 
calculated polarization of the Cu-O bilayer.
Although recent theories of FM-SC superlattices\cite{sipr,melo} 
are not strictly applicable to this atomic
scale SC/FM superlattice, the conditions necessary for interlayer FM
coupling\cite{melo} are present: the SC state must not be destroyed
by the proximity to the FM layer (the induced magnetization is small) and
the FM/SC interface roughness must be small (here it is atomically smooth).  
One likelihood is that the exchange
coupling will decrease below T$_S$ due to the SC gap,
which could be observable in the $q_z$ dependence of the spin waves.

We now summarize.  Our considerations show how coexistence of SC with FM
is possible: (i) the average magnetization is not large (1/30 that of iron,
in the case of Ru1212), (ii) the SC and FM subsystems are disjoint, in this
case precisely and thinly layered, (iii) both SC and FM layers are thin
enough to allow coupling perpendicular to the layers, hence three
dimensional ordering, and (iv) the
chemical bonding is such that coupling between the FM and SC layers is
weak enough (especially on one Fermi surface sheet) not to entirely disallow
superconductivity, yet strong enough to require an FFLO phase. 
RuSr$_2$GdCu$_2$O$_8$ presents a striking illustration of
behavior that can arise only in a sufficiently complex crystal structure
with several competing interactions.
 
We gratefully acknowledge close communication with J. Tallon and his
communication of unpublished work.  
This work was supported by Office of Naval Research
Grant No. N00014-97-1-0956 and National Science foundation Grant
DMR-9802076. 



\begin{references}
\bibitem{triplet}An alternative, p-wave pairing in the RuO$_2$
layer, is suggested by (1) the confirmed triplet SC in
Sr$_2$RuO$_4$ and (2) Ru1212 contains a substructure isomorphic
to Sr$_2$RuO$_4$.  However, the experimental evidence indicates 
cuprate superconductivity.
\bibitem{Gd}Whether the Gd moment is
aligned parallel or antiparallel to the Ru moment does not affect the
electronic structure in the CuO$_2$ or RuO$_2$ layers, reflecting
strongly localized Gd $f$ states.
\end{references}
\end{document}